\documentclass[twocolumn,10pt]{article}
\usepackage{amssymb,amsmath,epsfig}
\usepackage[dvips]{color}
\usepackage{graphicx}

\begin{document}
\title{\bf Anisotropic Multiverse with Varying $c$, $G$ and Study of Thermodynamics}

\author{\bf{Ujjal Debnath\thanks{ujjaldebnath@gmail.com} ~ and Soumak Nag\thanks{nagsoumak1996@gmail.com}}
 \\
\\
Department of Mathematics, Indian Institute of Engineering\\
Science and Technology, Shibpur, Howrah-711 103, India.}

\maketitle

\begin{abstract}
We assume the anisotropic model of the Universe in the framework
of varying speed of light $c$ and varying gravitational constant
$G$ theories and study different types of singularities. For the
singularity models, we write the scale factors in terms of cosmic
time and found some conditions for possible singularities. For
future singularities, we assume the forms of varying speed of
light and varying gravitational constant. For regularizing big
bang singularity, we assume two forms of scale factors: sine model
and tangent model. For both the models, we examine the validity of
null energy condition and strong energy condition. Start from the
first law of thermodynamics, we study the thermodynamic behaviours
of $n$ number of Universes (i.e., Multiverse) for (i) varying $c$,
(ii) varying $G$ and (iii) both varying $c$ and $G$ models. We
found the total entropies for all the cases in the anisotropic
Multiverse model. We also found the nature of the Multiverse if
total entropy is constant.
\end{abstract}


\section{Introduction}

The Multiverse theory states that there are multiple versions of
the universe, each of them are slightly different from the other.
Each universe is known as parallel Universe and the whole
collection of these parallel Universes is known as the Multiverse
(for review see \cite{Carr}). The notion of a cosmic Multiverse
has been introduced in the context of eternal inflation
\cite{Gibb,Vil} caused by inflaton field. Thus due to inflationary
expansion, the continuously growing number of subuniverses
produced which is surrounded by rapidly growing regions of
inflating space \cite{Reb}. In ref. \cite{Reb}, the model of
Multiverse has been discussed by taking scalar field model of dark
energy. Observational consequences of an interacting Multiverse
have been studied in \cite{Rob}. Marosek et al \cite{Mar} have
analyzed isotropic cyclic Multiverse model and studied several
kinds of singularities for varying speed of light $c$ and varying
gravitational constant $G$. Also they have studied the
thermodynamics for doubleverse model. The scenarios of parallel
cyclic Multiverses with varying $G$ have been studied in
\cite{Robl}. They have described the formalism of quantum
entanglement in the context of the Multiverse and thermodynamics
of entanglement for the cosmological models. The pre-inflation
scenario from the Multiverse has been studied in \cite{Mora}.
Canonically quantized Multiverse has been studied in\cite{Bou}.
The vacuum decay in an interacting Multiverse has been studied in
\cite{Roble}. Marosek et al \cite{Mar} have considered isotropic
cyclic Multiverse model and studied several kinds of singularities
for varying speed of light $c$ and varying gravitational constant
$G$ and regularizing the singularities in cyclic universe. Also
they have studied the thermodynamics for doubleverse model and
found the nature of the doubleverse if the total entropy of the
doubleverse is constant. Also the regularizing cosmological
singularities have been studied in \cite{Mar,Mar2,Mar3} by varying
physical constant (say, $c$, $G$) theories. Motivated by their
works, here we study the singularities in the anisotropic model of
the universe for varying $c$ and $G$. We study the regularizing
singularities for varying constants in anisotropic model of the
cyclic universe. We discuss the thermodynamic behaviors in
anisotropic Multiverse ($n$ number of universes) model and
investigate the nature of the Multiverse if total entropy of the
Multiverse is constant. The paper is organized as follows: In
section II, we study the anisotropic space-time model with varying
$c$ and $G$. In section III, we describe some kinds of
singularities. In section IV, we discuss regularizing big bang and
big rip singularities for cyclic model. In section V, we study the
thermodynamics in Multiverse with varying $c$ and $G$. The
discussions and some concluding remarks are presented in section
VI.

\section{Anisotropic Model of the Universe}

Marosek et al \cite{Mar} have studied isotropic model of the
universe for varying speed of light and varying gravitational
constant \cite{A1,B1,B2}. In this section, we study the
anisotropic model of the Universe and we write the Einstein's
field equations with continuity equation in varying speed of light
and varying gravitational constant. Now we consider homogeneous
and anisotropic space-time model of the Universe described by the
line element \cite{Chak1,Chak2,Chak3,UD2}
\begin{equation}
ds^{2}=-c^{2}(t)dt^{2}+a^{2}(t)dx^{2}+b^{2}(t)d\Omega_{k}^{2}
\end{equation}

where  $a$, $b$  are  functions  of  time  $t$ only and
\begin{eqnarray}d\Omega_{k}^{2}= \left\{\begin{array}{lll}
dy^{2}+dz^{2}, ~~~~~~~~~~~~ \text{when} ~~~k=0 ~~~~\\ ~~~~~~~~~~~~~~~~~~~( \text{Bianchi ~I ~model})\\
d\theta^{2}+sin^{2}\theta d\phi^{2}, ~~~~~ \text{when}
~~~k=+1~~\\~~~~~~~~~~~~~~~~~~
( \text{Kantowaski-Sachs~ model})\\
d\theta^{2}+sinh^{2}\theta d\phi^{2}, ~~~ \text{when} ~~~k=-1
~~\\~~~~~~~~~~~~~~~~~ (\text{Bianchi~ III~ model})
\end{array}\right.
\end{eqnarray}

Here  $k~(=0,+1,-1)$  is  the  curvature  index  of  the
corresponding 2-space, so  that  the  above  three  types  are
described  by Thorne \cite{Thorne}  as  flat, closed  and  open
respectively and $c=c(t)$ is the time varying speed of light.\\

As per the proposal of varying speed of light \cite{AM}, the
Friedman equations remain valid even when $\dot{c}\ne 0$. Due to
the same prescription, the Einstein's field equations in the
framework of varying speed of light and varying gravitational
constant theories in anisotropic universe are
\cite{Chak1,Chak2,Chak3}
\begin{equation}
\frac{\dot{b}^{2}}{b^{2}}
+2\frac{\dot{a}}{a}\frac{\dot{b}}{b}+\frac{kc^{2}(t)}{b^{2}}=8\pi
G(t) \rho~,
\end{equation}
\begin{equation}
\frac{2\ddot{b}}{b}+\frac{\dot{b}^{2}}{b^{2}}+\frac{kc^{2}(t)}{b^{2}}=-\frac{8\pi
G(t)p}{c^{2}(t)}~,
\end{equation}
\begin{equation}
\frac{\ddot{a}}{a}+\frac{\ddot{b}}{b}+\frac{\dot{a}}{a}\frac{\dot{b}}{b}=-\frac{8\pi
G(t)p}{c^{2}(t)}
\end{equation}
where $\rho$ and $p$ are respectively the mass density and
pressure. Consequently, $\rho c^{2}$ is the energy density. Here
$G=G(t)$ is the time varying gravitational constant.\\

From equations (3) - (5), we obtain
\begin{equation}
\frac{\ddot{a}}{a}+2\frac{\ddot{b}}{b}=-4\pi
G(t)\left(\rho+\frac{3p}{c^{2}(t)} \right)
\end{equation}
For varying $c$ and varying $G$, the modified continuity equation
is obtained as
\begin{equation}
\dot{\rho}+\left(\frac{\dot{a}}{a}+2\frac{\dot{b}}{b}\right)\left(\rho+\frac{p}{c^{2}}
\right)=-\rho\frac{\dot{G}}{G}+\frac{kc\dot{c}}{4\pi Gb^{2}}
\end{equation}
In the next sections, we study the singularities for varying
constants in anisotropic space-time model.

\section{Singularities}

In this section, we study several singularities in the anisotropic
Universe. For varying $c$ and varying $G$ models, we find the
conditions so that singularities can occur. For two scale factors
model (anisotropic), big bang initial singularity \cite{Tipler}
occurs at $t\rightarrow 0$ and at this time, $a\rightarrow  0$,
$b\rightarrow  0$, $\rho\rightarrow \infty$ and $|p|\rightarrow
\infty$. The future singularities can be
classified in the following ways:\\

$\bullet$ Type I (Big Rip) \cite{Cald}: For $t\rightarrow t_s$,
$a\rightarrow \infty$, $b\rightarrow \infty$, $\rho\rightarrow \infty$ and $|p|\rightarrow \infty$.\\

$\bullet$ Type II (Sudden future) \cite{Barr}: For $t\rightarrow
t_s$, $a\rightarrow a_s$, $b\rightarrow b_{s}$, $\rho\rightarrow \rho_s$ and $|p|\rightarrow \infty$.\\

$\bullet$ Type III (Finite scale factors) \cite{Nojiri}:
$t\rightarrow t_s$, $a\rightarrow a_s$, $b\rightarrow b_{s}$, $\rho\rightarrow \infty$ and $|p|\rightarrow \infty$.\\

$\bullet$ Type IV (Big separation) \cite{Nojiri}: For
$t\rightarrow t_s$, $a\rightarrow a_s$, $b\rightarrow b_{s}$, $\rho\rightarrow 0$ and $|p|\rightarrow 0$.\\

$\bullet$ Type V ($w$-singularity) \cite{Dab}: For $t\rightarrow
t_s$,
$a\rightarrow a_s$, $b\rightarrow b_{s}$, $\rho\rightarrow 0$, $|p|\rightarrow 0$, $w=p/\rho\rightarrow\infty$.\\

$\bullet$ Type VI (Little Rip) \cite{Fram}: For $t\rightarrow
\infty$, $a\rightarrow \infty$, $b\rightarrow \infty$, $\rho\rightarrow \infty$ and $|p|\rightarrow \infty$.\\

where $t_s$, $a_s$, $b_{s}$ and $\rho_s$ are constants with
$a_s\neq 0$, $b_{s}\neq 0$. The big bang, big rip and little rip
are strong singularities while the other singularities mentioned
above are weak singularities. Since $a(t)$ and $b(t)$ are unknown
functions of $t$, so to keep anisotropy of the Universe, we must
have $a(t)/b(t)\ne$ constant. So for simplicity of calculation, we
may assume, $b(t)$ is related to the power law form of $a(t)$
i.e., $b(t)=b_{0}a^{\alpha}(t)$
\cite{Chak1,Chak2,Chak3,UD2,UD4,UD5,UD6,UD7,UD8}, where
$b_{0},\alpha$ are positive constants. The physical importance of
this assumption is that it gives constant ratio of shear and
expansion scalar \cite{a1}. Here we use the two scale factors
which after appropriate choice of parameters admit big-bang,
big-rip, sudden future, finite scale factor and $w$-singularities
and read as \cite{Mar,Mar2}
\begin{equation}
a(t)=a_{s}\left(\frac{t}{t_{s}}\right)^{m}~exp\left|1-\frac{t}{t_{s}}
\right|^{n}
\end{equation}
where $a_{s},t_{s},m,n$ are positive constants. So $b(t)$ can be
written as
\begin{equation}
b(t)=b_{0}a^{\alpha}_{s}\left(\frac{t}{t_{s}}\right)^{m\alpha}~exp\left(\alpha\left|1-\frac{t}{t_{s}}
\right|^{n}\right)
\end{equation}
Since first powers of equations (8) and (9) give the big bang
singularity, so the second parts of (8) and (9) provide the future
singularity. Now we want to analyze the future singularity for our
anisotropic model. For future singularity part, taking the
expressions of $a$, $b$ from equations (8), (9) and putting their
values in equations (3) and (4), we directly obtain the energy
density and pressure for the future singularity as \cite{Mar,Mar2}
\begin{eqnarray}
\rho=\frac{1}{8\pi
G(t)}\left[\frac{(\alpha^{2}+2\alpha)n^{2}}{t_{s}^{2}}\left|1-\frac{t}{t_{s}}
\right|^{2n-2} \right. \nonumber\\
\left.
+\frac{kc^{2}(t)}{a_{s}^{2\alpha}}~exp\left(-\alpha\left|1-\frac{t}{t_{s}}
\right|^{n}\right) \right]
\end{eqnarray}
and
\begin{eqnarray}
p=-\frac{c^{2}(t)}{12\pi
G(t)}\left[\frac{(5\alpha^{2}+2\alpha+2)n^{2}}{2t_{s}^{2}}\left|1-\frac{t}{t_{s}}
\right|^{2n-2} \right. \nonumber\\
+\frac{(1+2\alpha)n(n-1)}{t_{s}^{2}}\left|1-\frac{t}{t_{s}}
\right|^{n-2}  \nonumber\\
\left.
+\frac{kc^{2}(t)}{2a_{s}^{2\alpha}}~exp\left(-\alpha\left|1-\frac{t}{t_{s}}
\right|^{n}\right) \right]
\end{eqnarray}
We observe that for $0 \le n \le 1$, there is a finite scale
factor singularity, for $1\le n \le 2$, there is a sudden future
singularity and for $n \ge 2$, there is a generalized sudden
future singularity. Since future singularity occurs at $t=t_{s}$,
so we may assume the variable $G$ is in the power law form
\cite{Mar,Mar2}:
\begin{equation}
G(t)=G_{0}\left|1-\frac{t}{t_{s}} \right|^{-r}
\end{equation}
where $r>0$ and $G_{0}>0$ are constants. For this form of $G(t)$,
at $t=t_{s}$, we have $G\rightarrow\infty$. At $t=t_{s}$ the
density and pressure are finite for $r > 2-n$. So sudden
singularity is regularized due to strong gravitational coupling
($G\rightarrow\infty$ at $t=t_{s}$). But for $t\rightarrow\infty$,
the scale factors $a\rightarrow\infty$ and $b\rightarrow\infty$
and both the density $\rho \rightarrow\infty$ and pressure
$|p|\rightarrow\infty$ which achieved to a little rip singularity.
On the other hand, similar to the choice of $G(t)$, we may choose
the varying speed of light $c$ as in the power law form
\cite{Mar,Mar2} :
\begin{equation}
c(t)=c_{0}\left|1-\frac{t}{t_{s}} \right|^{\frac{\beta}{2}}
\end{equation}
where $\beta>0$ and $c_{0}>0$ are constants. For this form of
$c(t)$, at $t=t_{s}$, we have $c\rightarrow 0$. But for $\beta
> 2-n$ at $t = t_{s}$, $a\rightarrow\infty$ and
$b\rightarrow\infty$ and both the density $\rho \rightarrow\infty$
and pressure $p\rightarrow\infty$ which regularized to a sudden
singularity. Also for $t\rightarrow\infty$, we obtain
$a\rightarrow\infty$, $b\rightarrow\infty$,
$\rho\rightarrow\infty$ and $|p|\rightarrow\infty$ which achieved
at little rip singularity. So due to above choices of the varying
$c$ and $G$, sudden singularity can be regularized.

\section{Regularizing Strong Singularity with Varying $G$}

In this section, we study regularizing strong singularities like
big bang and big rip singularities for varying $G$ model in
anisotropic cyclic universe. So we assume constant $c$ in order to
construct regularized anisotropic universe \cite{Mar}. In the
following subsection A, we discuss the sine model to study
regularizing big bang singularity while in the subsection B, we
discuss the tangent model to study regularizing  big bang and big
rip singularities. Also we study the validity of the energy
conditions for both models.

\subsection{Regularizing big bang singularity: sine model}

Here for cyclic universe, we assume the `sine' model to describe
regularizing big bang singularity. So the scale factors $a(t)$ and
$b(t)$ can be chosen in the forms \cite{Mar}:
\begin{equation}
a(t)=a_{0}\left|sin\left(\pi \frac{t}{t_{c}}\right) \right|
\end{equation}
and
\begin{equation}
b(t)=b_{0}a^{\alpha}(t)=b_{0}a_{0}^{\alpha}\left|sin^{\alpha}\left(\pi
\frac{t}{t_{c}}\right) \right|
\end{equation}
where $a_{0},b_{0},\alpha$ are positive constants. Here $t_{c}$ is
the turning point for the cyclic universe where scale factors $a$
and $b$ are zero. For $t\rightarrow 0$ we get $a\rightarrow 0$ and
$b\rightarrow 0$, which follows big bang singularity. For constant
$c$, we assume
\begin{equation}
G(t)=\frac{G_{0}}{ab}
\end{equation}
where $G_{0}$ is positive constant. So at the big bang singularity
(at $t=0$), $G\rightarrow\infty$. But at the turning point
($t=t_{c}$), we have $a\rightarrow 0$, $b\rightarrow 0$ and
$G\rightarrow\infty$. That means at the turning point of the
cyclic universe, the curvature singularity (big-bang like
singularity) can be regularized due to strong gravitational
coupling. Here `like' is used because the density and pressure are
regular for this singularity but they are not regular at a big
bang. Using Einstein's field equations, we obtain the density and
pressure in the forms:
\begin{eqnarray}
\rho=\frac{a_{0}^{2\alpha+1}b_{0}^{2}}{8\pi
G_{0}}~sin^{2\alpha+1}\left(\pi
\frac{t}{t_{c}}\right)\left[(\alpha^{2}+2\alpha)cot^{2}\left(\pi
\frac{t}{t_{c}}\right) \right.\nonumber\\
\left.
+\frac{kc^{2}}{a_{0}^{2\alpha}b_{0}^{2}}~cosec^{2\alpha}\left(\pi
\frac{t}{t_{c}}\right) \right]
\end{eqnarray}
and
\begin{eqnarray}
p=\frac{\pi c^{2}}{12G_{0} t_{c}^{2}}
\left[(1+2\alpha)-3\alpha^{2}cot^{2}\left(\pi
\frac{t}{t_{c}}\right)\right] \nonumber\\
-\frac{k}{24\pi G
a_{0}^{2\alpha}b_{0}^{2}}~cosec^{2\alpha}\left(\pi
\frac{t}{t_{c}}\right)
\end{eqnarray}
Now we study null energy condition and strong energy condition for
this model. From the above expressions (17) and (18), we obtain
\begin{eqnarray}
p+\rho c^{2}=\frac{\pi c^{2}a_{0}^{\alpha+1}b_{0}}{12
G_{0}t_{c}^{2}}~ \left|sin^{\alpha+1}\left(\pi
\frac{t}{t_{c}}\right)\right| \times \nonumber\\
\left[(1+2\alpha)+(4\alpha-\alpha^{2})cot^{2}\left(\pi
\frac{t}{t_{c}}\right)\right]  \nonumber\\
+\frac{kc^{4}}{12a_{0}^{\alpha-1}b_{0}}~cosec^{\alpha}\left(\pi
\frac{t}{t_{c}}\right)
\end{eqnarray}
and
\begin{eqnarray}
3p+\rho c^{2}=\frac{\pi c^{2}a_{0}^{\alpha+1}b_{0}}{4
G_{0}t_{c}^{2}}~\left|sin^{\alpha+1}\left(\pi
\frac{t}{t_{c}}\right)\right| \times \nonumber\\
\left[(1+2\alpha)+2(\alpha-\alpha^{2})cot^{2}\left(\pi
\frac{t}{t_{c}}\right)\right]
\end{eqnarray}
The null energy condition $p+\rho c^{2}>0$ is satisfied for
$0<\alpha\le 4$ with $k\ge 0$. The strong energy condition $3p+\rho c^{2}>0$ is satisfied for $0<\alpha\le 1$.\\

At the turning points, we have\\
\begin{eqnarray}
a(nt_{c})=a_{0}~,~~b(nt_{c})=b_{0}a_{0}^{\alpha}~,~G(nt_{c})=\frac{G_{0}}{b_{0}a_{0}^{\alpha+1}}~,\nonumber\\
~\rho(nt_{c})=\frac{kc^{2}}{8\pi G_{0} b_{0}a_{0}^{\alpha-1}}
\end{eqnarray}
and
\begin{eqnarray}
p(nt_{c})=\frac{\pi c^{2}b_{0}a_{0}^{\alpha+1}}{12G_{0}
t_{c}^{2}}~(1+2\alpha)-\frac{kc^{4}}{24\pi G_{0}
b_{0}a_{0}^{\alpha-1}}
\nonumber\\
=-\frac{1}{3}~c^{2}\rho(nt_{c}) +\frac{\pi
c^{2}b_{0}a_{0}^{\alpha+1}}{12G_{0} t_{c}^{2}}~(1+2\alpha)
\end{eqnarray}
where $n=\frac{1}{2},~\frac{3}{2},~\frac{5}{2},...$. But at
$t=mt_{c}$ ($m=0,1,2,...$), $a\rightarrow 0$, $b\rightarrow 0$,
$\rho=$ constant, $p=$ constant and $G\rightarrow\infty$. So the
curvature singularity (big-bang like singularity) is
regularized in each cycle due to strong gravitational coupling.\\

Now the Hubble parameter is defined by
\begin{equation}
H=\frac{1}{3}\left(\frac{\dot{a}}{a}+2\frac{\dot{b}}{b}
\right)=\frac{\pi}{3t_{c}}(1+2\alpha)~cot\left(\pi\frac{t}{t_{c}}
\right)
\end{equation}
and its derivative is obtained by
\begin{equation}
\dot{H}=-\frac{\pi^{2}}{3t_{c}^{2}}(1+2\alpha)~cosec^{2}\left(\pi\frac{t}{t_{c}}
\right)
\end{equation}
We observe that for $mt_{c}\le t \le (m+1)t_{c}$, $H>0$ which
represents the expansion and for $(m+1)t_{c}\le t \le (m+2)t_{c}$,
$H<0$ which represents the contraction. Also since $\dot{H}<0$, so
there is no proper bounce of the universe.

\subsection{Regularizing big bang and big rip singularities: tangent model}

Here we consider `tangent' model to describe regularizing big bang
and big rip singularities. So the forms of the scale factors are
\cite{Mar}
\begin{equation}
a(t)=a_{0}\left|tan\left(\frac{\pi t}{t_{s}}\right) \right|~,
\end{equation}
and
\begin{equation}
b(t)=b_{0}a^{\alpha}(t)=b_{0}a^{\alpha}_{0}\left|tan^{\alpha}\left(\frac{\pi
t}{t_{s}}\right) \right|
\end{equation}
and gravitational constant as
\begin{equation}
G(t)=\frac{G_{s}}{b^{\beta}(t)}=\frac{G_{s}}{b^{\beta}_{0}a^{\alpha\beta}_{0}
\left|tan^{\alpha\beta}\left(\frac{\pi t}{t_{s}}\right)\right|}
\end{equation}
The mass density and pressure are obtained by
\begin{eqnarray}
\rho=\frac{b^{\beta}_{0}a^{\alpha\beta}_{0}}{8\pi G_{s}t_{s}^{2}}
\left|tan^{\alpha\beta}\left(\frac{\pi
t}{t_{s}}\right)\right|\left[4\pi^{2}(\alpha^{2}+2\alpha)
~cosec^{2}\left(2\frac{\pi t}{t_{s}}\right) \right. \nonumber\\
\left. +\frac{kc^{2}t_{s}^{2}}{b_{0}^{2}a_{0}^{2\alpha}}
~\left|cot^{2\alpha}\left(\frac{\pi t}{t_{s}}\right)\right|
\right]~~~~~~~~
\end{eqnarray}
and
\begin{eqnarray}
p=\frac{\pi c^{2}b^{\beta}_{0}a^{\alpha\beta}_{0}}{6
G_{s}t_{s}^{2}} \left|tan^{\alpha\beta}\left(\frac{\pi
t}{t_{s}}\right)\right|
\times \nonumber\\
\left[-(5\alpha^{2}+2\alpha+2)
~cosec^{2}\left(2\frac{\pi t}{t_{s}}\right) \right. \nonumber\\
 +2(1+2\alpha)
~cosec\left(2\frac{\pi t}{t_{s}}\right)cot\left(2\frac{\pi
t}{t_{s}}\right)
\nonumber\\
\left. -\frac{kc^{2}t_{s}^{2}}{4\pi^{2}b_{0}^{2}a_{0}^{2\alpha}}
~\left|cot^{2\alpha}\left(\frac{\pi t}{t_{s}}\right)\right|
\right]
\end{eqnarray}
The minimum values of the mass density and pressure are given by
\begin{equation}
\rho_{min}=\rho(\kappa
t_{s})=\frac{b^{\beta}_{0}a^{\alpha\beta}_{0}}{8\pi
G_{s}t_{s}^{2}}\left[4\pi^{2}(\alpha^{2}+2\alpha)+\frac{kc^{2}t_{s}^{2}}{b_{0}^{2}a_{0}^{2\alpha}}
\right]
\end{equation}
and
\begin{equation}
p_{min}=p(\kappa
t_{s})=-\frac{c^{2}b^{\beta}_{0}a^{\alpha\beta}_{0}}{24\pi
G_{s}t_{s}^{2}}\left[4\pi^{2}(5\alpha^{2}+2\alpha+2)+\frac{kc^{2}t_{s}^{2}}{b_{0}^{2}a_{0}^{2\alpha}}
\right]
\end{equation}
where $\kappa=\frac{2m+1}{4}~(m=0,1,2,...)$.\\

We see that the scale factors $a\rightarrow\infty$,
$b\rightarrow\infty$, $G\rightarrow 0$, density $\rho=$ constant
and pressure $p=$ constant for $t=nt_{s}$ with
$n=1/2,3/2,5/2,...$. So the big rip like singularity is
regularized. But the scale factors $a\rightarrow 0$, $b\rightarrow
0$, $G\rightarrow \infty$, density $\rho=$ constant and pressure
$p=$ constant for $t=mt_{s}$ with
$m=0,1,2,....$. So the curvature singularity (big-bang like singularity) is also regularized. \\

Now we study null energy condition and strong energy condition for
this model. Now from equations (28) and (29), we get
\begin{eqnarray}
p+\rho c^{2}=\frac{\pi c^{2}b^{\beta}_{0}a^{\alpha\beta}_{0}}{3
G_{s}t_{s}^{2}} \left|tan^{\alpha\beta}\left(\frac{\pi
t}{t_{s}}\right)\right|
\times \nonumber\\
\left[-(\alpha^{2}+1)
~cosec^{2}\left(2\frac{\pi t}{t_{s}}\right) \right. \nonumber\\
 +(1+2\alpha)
~cosec\left(2\frac{\pi t}{t_{s}}\right)cot\left(2\frac{\pi
t}{t_{s}}\right)
\nonumber\\
\left. +\frac{kc^{2}t_{s}^{2}}{4\pi^{2}b_{0}^{2}a_{0}^{2\alpha}}
~\left|cot^{2\alpha}\left(\frac{\pi t}{t_{s}}\right)\right|
\right]
\end{eqnarray}

At $t=mt_{s}$, we get: (i) $p+\rho c^{2}>0$ i.e., null energy
condition is satisfied for $0<\alpha\le 2$ and $k\ge 0$; (ii)
$p+\rho c^{2}<0$ i.e., null energy condition is violated for
$\alpha>2$ and $k<0$. But at $t=nt_{s}$, we get: $p+\rho c^{2}<0$
i.e., null energy condition is always violated.\\

Also from equations (28) and (29), we have
\begin{eqnarray}
3p+\rho c^{2}=\frac{\pi c^{2}b^{\beta}_{0}a^{\alpha\beta}_{0}}{
G_{s}t_{s}^{2}} \left|tan^{\alpha\beta}\left(\frac{\pi
t}{t_{s}}\right)\right|
\times\nonumber\\
\left[-(2\alpha^{2}+1) ~cosec^{2}\left(2\frac{\pi t}{t_{s}}\right)
\right. \nonumber\\
\left. +(1+2\alpha) ~cosec\left(2\frac{\pi
t}{t_{s}}\right)cot\left(2\frac{\pi t}{t_{s}}\right) \right]
\end{eqnarray}

At $t=mt_{s}$, we get: (i) $3p+\rho c^{2}>0$ i.e., strong energy
condition is satisfied for $0<\alpha\le 1$; (ii) $3p+\rho c^{2}<0$
i.e., strong energy condition is violated for $\alpha>1$. But at
$t=nt_{s}$, we get: $3p+\rho c^{2}<0$
i.e., strong energy condition is always violated.\\

\section{Thermodynamics in Multiverse}

In ref. \cite{Mar}, the authors have studied the classical
thermodynamics of two universes (doubleverse) and their
consideration that the entropy of the two universes is changing in
such correlated way that the total entropy is always the same
(constant). Motivated by their work, in this section, we will
study the thermodynamics in Multiverse ($n$ number of universes)
model. For this purpose, we use the first law of thermodynamics
and calculate the total entropy of the Multiverse for (i) varying
$c$ with constant $G$ model, (ii) varying $G$ with constant $c$
model and (iii) both varying $c$ and varying $G$ model. Then we'll
show that the total entropy is always the same (constant) provided
there are some relations between $n$ number of universes.

\subsection{Thermodynamics for varying $c$}

First we consider the varying $c$ model with constant $G$ and
analyze the thermodynamic nature of $n$ number of universes i.e.,
Multiverse model. The first law of thermodynamics is given by
\cite{Mar}
\begin{equation}
dE=TdS-pdV
\end{equation}
where $E,T,S,p,V$ are the internal energy, temperature, entropy,
pressure and volume respectively. From Einstein's mass-energy
equation we have
\begin{equation}
E=mc^{2}=\rho Vc^{2}
\end{equation}
where $\rho$ is the mass density and $V=ab^{2}$. From above two
equations, we obtain \cite{Mar}
\begin{equation}
\dot{\rho}+\frac{\dot{V}}{V}\left(\rho+\frac{p}{c^{2}}
\right)+2\rho\frac{\dot{c}}{c}-\frac{T}{Vc^{2}}\dot{S}=0
\end{equation}
Using continuity equation
\begin{equation}
\dot{\rho}+\left(\frac{\dot{a}}{a}+2\frac{\dot{b}}{b}\right)\left(\rho+\frac{p}{c^{2}}
\right)=\frac{kc\dot{c}}{4\pi Gb^{2}}
\end{equation}
and above equation (36), we have
\begin{equation}
\frac{T}{Vc^{2}}\dot{S}-2\rho\frac{\dot{c}}{c}=\frac{kc\dot{c}}{4\pi
Gb^{2}}
\end{equation}
Defining \cite{Mar}
\begin{equation}
\tilde{\rho}=\frac{1}{8\pi G}\left( \frac{\dot{b}^{2}}{b^{2}}
+2\frac{\dot{a}}{a}\frac{\dot{b}}{b}+\frac{2kc^{2}}{b^{2}}\right)=\rho+\frac{kc^{2}}{8\pi
Gb^{2}}
\end{equation}
and
\begin{equation}
\tilde{p}=p-\frac{kc^{2}}{24\pi Gb^{2}}
\end{equation}
we obtain
\begin{equation}
\dot{S}=2\tilde{\rho}\frac{Vc^{2}}{T}\frac{\dot{c}}{c}
\end{equation}
For the equation of state of ideal gas we can take \cite{Mar}
\begin{equation}
\tilde{\rho}\frac{Vc^{2}}{T}=constant=\frac{Nk_{B}}{\tilde{w}}
\end{equation}
where $N$ is the number of particles, $k_{B}$ is the Boltzmann
constant and $\tilde{w}=\frac{\tilde{p}}{\tilde{\rho}c^{2}}$.
After integrating the above equation, we get the entropy as in the
following form:
\begin{equation}
S(t)=\frac{2Nk_{B}}{\tilde{w}}~log[A_{0}c(t)]
\end{equation}
where $A_{0}$ is constant of integration. For flat ($k = 0$)
model, we have $\tilde{\rho}=\rho$. Now using the barotropic
equation of state $p=w\rho c^{2}$ with the equation of state
parameter $w = $ const, we have
\begin{equation}
S(t)=\frac{2Nk_{B}}{w}~log[A_{0}c(t)]
\end{equation}
The nature of entropy depends of the nature of $c(t)$.\\

Now we want to study the nature of entropy in the Multiverse
model. So the total entropy of the $n$ number of universes (i.e.,
Multiverse) is given by
\begin{equation}
\dot{S}=\sum_{i=1}^{n}\dot{S}_{i}(t)
\end{equation}
where the entropy of $i$-th universe is
\begin{equation}
S_{i}(t)=\frac{2N_{i}k_{B}}{\tilde{w}}~log[A_{0}c_{i}(t)]~,
~i=1,2,...,n
\end{equation}
We assume the following form of $c(t)$ \cite{Mar}:
\begin{equation}
A_{0}c_{i}(t)=e^{\lambda_{i}\phi_{i}(t)}~, ~i=1,2,...,n
\end{equation}
where $\lambda_{i}$'s are arbitrary constants. So the total
entropy of the Multiverse is \cite{Mar}
\begin{equation}
S=\sum_{i=1}^{n}S_{i}(t)=\sum_{i=1}^{n}\frac{2k_{B}N_{i}\lambda_{i}}{\tilde{w}}~\phi_{i}(t)
\end{equation}
Now we can study the relation between the natures of the $n$
number of universes if we assume total entropy of the Multiverse
is constant. In ref \cite{Mar}, the authors have considered the
doubleverse system and they have studied the nature of two
universes if the total entropy of two universes is constant. For
this purpose, here we may assume in the Multiverse there are even
number of universes (i.e., $n$ is even). To get constant total
entropy of the Multiverse (i.e., $S=$ constant), we may assume
\cite{Mar}
\begin{equation}
\phi_{i}(t)=\left\{
\begin{array}{l}
sin^{2}\left(\frac{\pi
t}{t_{s}}\right)~,~i=1,2,...,\frac{n}{2}\\\\
cos^{2}\left(\frac{\pi t}{t_{s}}\right)~,~i=\frac{n}{2}+1,
\frac{n}{2}+2,...,n
\end{array}
\right.
\end{equation}
with
\begin{equation}
N_{1}\lambda_{1}=N_{2}\lambda_{2}=...=N_{n}\lambda_{n}
\end{equation}
From these, we may conclude that if the entropies of $n/2$ number
of universes are growing/diminishing and entropies of other $n/2$
number of universes are diminishing/growing then total entropy of
the Multiverse may be constant. In this case, the speed of light
of $n/2$ number of universes are growing/diminishing and it's
value of other $n/2$ number of universes are diminishing/growing.

\subsection{Thermodynamics for varying $G$}

Now we consider the varying $G$ model with constant $c$ and
analyze the thermodynamic nature of $n$ number of universes (i.e.,
Multiverse) model. For varying gravitational constant $G$ with
constant $c$, the continuity equation is
\begin{equation}
\dot{\rho}+\left(\frac{\dot{a}}{a}+2\frac{\dot{b}}{b}\right)\left(\rho+\frac{p}{c^{2}}
\right)=-\rho\frac{\dot{G}}{G}
\end{equation}
So using equation (36), we get
\begin{equation}
\dot{S}=-\frac{\rho Vc^{2}}{T}\frac{\dot{G}}{G}
\end{equation}
Using the equation of state of ideal gas
\begin{equation}
\frac{\rho Vc^{2}}{T}=constant=Nk_{B}
\end{equation}
we get
\begin{equation}
S=Nk_{B}~log\left[\frac{B_{0}}{G(t)}\right]
\end{equation}
where $B_{0}$ is an integration constant.\\

So the entropy of $i$-th universe is
\begin{equation}
S_{i}(t)=N_{i}k_{B}~log\left(\frac{B_{0}}{G_{i}(t)}\right),
~i=1,2,...,n
\end{equation}
So the total entropy of the Multiverse is
\begin{equation}
S=\sum_{i=1}^{n}S_{i}=\sum_{i=1}^{n}N_{i}k_{B}~log\left(\frac{B_{i}}{G_{i}(t)}\right)
\end{equation}
Similar to the previous subsection $A$, we assume that the
Multiverse contains even number of universes (i.e., $n$ is even).
Now we assume the gravitational constant $G(t)$ in terms of the
scale factors as in the following form:
\begin{equation}
G_{i}(t)=\left\{
\begin{array}{l}
\frac{G_{0i}}{a_{i}(t)b_{i}(t)}~,~i=1,2,...,\frac{n}{2}\\\\
G_{0i}a_{i}(t)b_{i}(t)~,~i=\frac{n}{2}+1, \frac{n}{2}+2,...,n
\end{array}
\right.
\end{equation}
If all the scale factors of all the universes are equal i.e.,
$a_{i}(t)=a(t)$ and $b_{i}(t)=b(t),~i=1,2,...,n$ then all the
universes have same types of evolution. To get the constant total
entropy, we need the following conditions:
\begin{eqnarray}
N_{1}=N_{2}=...=N_{n},~~~\text{with}~~~
\frac{B_{1}}{G_{01}}=\frac{B_{2}}{G_{02}}=...=\frac{B_{n}}{G_{0n}}
\end{eqnarray}
We observe that the entropies and gravitational constant of $n/2$
number of universes are growing (or diminishing) and other $n/2$
number of universes are diminishing (or growing) but the total
entropy of the Multiverse will be constant.

\subsection{Thermodynamics for varying $c$ and $G$}

Finally, we consider both varying $c$ and varying $G$ models and
analyze the thermodynamic nature of $n$ number of universes i.e.,
Multiverse model. For varying $c$ and varying $G$ together, the
continuity equation becomes
\begin{equation}
\dot{\rho}+\left(\frac{\dot{a}}{a}+2\frac{\dot{b}}{b}\right)\left(\rho+\frac{p}{c^{2}}
\right)=-\rho\frac{\dot{G}}{G}+\frac{kc\dot{c}}{4\pi Gb^{2}}
\end{equation}
Using this equation and equation (36), we get
\begin{equation}
\frac{T}{Vc^{2}}\dot{S}-2\rho\frac{\dot{c}}{c}=-\rho\frac{\dot{G}}{G}+\frac{kc\dot{c}}{4\pi
Gb^{2}}
\end{equation}
Now we obtain
\begin{equation}\label{1b}
\dot{S}=2\frac{\tilde{\rho} Vc^{2}}{T}\frac{\dot{c}}{c}-\frac{\rho
Vc^{2}}{T}\frac{\dot{G}}{G}
\end{equation}
where $\tilde{\rho}$ is defined in equation (39). Due to ideal gas
equation of states, we may take
\begin{eqnarray}
\tilde{\rho}\frac{Vc^{2}}{T}=constant=\frac{Nk_{B}}{\tilde{w}},
\nonumber\\~~~and ~~~\rho\frac{Vc^{2}}{T}=constant=Nk_{B}
\end{eqnarray}
Using the above relations, integrating (\ref{1b}), we get the
entropy
\begin{equation}
S(t)=\frac{2Nk_{B}}{\tilde{w}}~log[c(t)]-Nk_{B}~log[G(t)]+log[D_{0}]
\end{equation}
where $D_{0}$ is an integration constant. The entropy of $i$-th
Universe is
\begin{eqnarray}
S_{i}(t)=\frac{2N_{i}k_{B}}{\tilde{w}}~log[c_{i}(t)]-N_{i}k_{B}~log[G_{i}(t)]+log[D_{i}]~,\nonumber\\
i=1,2,...,n ~~~~~
\end{eqnarray}
So the total entropy of the Multiverse is obtained as
\begin{eqnarray}
S=\sum_{i=1}^{n}S_{i}=&&\sum_{i=1}^{n}\left[\frac{2N_{i}k_{B}}{\tilde{w}}~log[c_{i}(t)]
\right. \nonumber\\ &&\left.
-N_{i}k_{B}~log[G_{i}(t)]+log[D_{i}]\right]
\end{eqnarray}
Since here both $c_{i}(t)$ and $G_{i}(t)$ are varying with $t$, so
to get constant total entropy of the $n$ numbers of universes
(where $n$ is considered as even) we can choose $c_{i}(t)$ from
equation (47) where $\phi_{i}(t)$ is satisfying the equation (49)
with the condition (50) and $G_{i}(t)$ is satisfying the equation
(57) with the condition (58). In this case, the speed of light and
gravitational constant of $n/2$ number of universes are
growing/diminishing and their values of other $n/2$ number of
universes are diminishing/growing. So for both varying speed of
light and varying gravitational constant, we
may conclude that total entropy of the Multiverse is constant.  \\

\section{Discussions and Concluding Remarks}

We have assumed the anisotropic model of the Universe in the
framework of varying speed of light $c$ and varying gravitational
constant $G$ theories. We have mentioned different types of weak
and strong singularities for the anisotropic model of the
Universe. To study the strong and weak singularity models, we have
written the scale factors $a(t)$ and $b(t)$ in terms of cosmic
time and found some conditions for possible weak and strong
singularities. For future singularity, the density and the
pressure have been obtained. We have observed that there is a
finite scale factor singularity if $0 \le n \le 1$. But there is a
sudden future singularity if $1\le n \le 2$. On the other hand,
there is a generalized sudden future singularity if $n \ge 2$. For
future singularity, we have assumed particular power law forms of
$c(t)$ and $G(t)$. For the choice of $G(t)$, at $t=t_{s}$ (the
future singularity time), we have obtained $G\rightarrow\infty$.
At $t=t_{s}$ the density and pressure are finite for $r > 2-n$. So
sudden singularity is regularized due to strong gravitational
coupling ($G\rightarrow\infty$ at $t=t_{s}$). But for
$t\rightarrow\infty$, the scale factors $a\rightarrow\infty$ and
$b\rightarrow\infty$ and both the density $\rho \rightarrow\infty$
and pressure $|p|\rightarrow\infty$ which achieved to a little rip
singularity. For the choice of $c(t)$, at $t=t_{s}$, we have
obtained $c\rightarrow 0$. But for $\beta
> 2-n$ at $t = t_{s}$, $a\rightarrow\infty$ and
$b\rightarrow\infty$, $\rho \rightarrow\infty$ and
$p\rightarrow\infty$ which regularized to a sudden singularity.
Also for $t\rightarrow\infty$, we have obtained
$a\rightarrow\infty$, $b\rightarrow\infty$,
$\rho\rightarrow\infty$ and $|p|\rightarrow\infty$ which achieved
at little rip singularity.\\

For strong singularity with varying $G$, we have assumed two forms
of scale factors: sine model and tangent model. For sine model, at
the big bang singularity (at $t=0$), we have $G\rightarrow\infty$.
Also at the turning point $t=mt_{c}$ ($m=0,1,2,...$), we have
obtained $a\rightarrow 0$, $b\rightarrow 0$, $\rho=$ constant,
$p=$ constant and $G\rightarrow\infty$. That means at the turning
point of the cyclic universe, the big bang like singularity is
regularized due to strong gravitational coupling. But for tangent
model, the scale factors $a\rightarrow\infty$,
$b\rightarrow\infty$, $G\rightarrow 0$, $\rho=$ constant and $p=$
constant for $t=nt_{s}$ with $n=1/2,3/2,5/2,...$. So the big rip
like singularity is regularized. But the scale factors
$a\rightarrow 0$, $b\rightarrow 0$, $G\rightarrow \infty$, density
$\rho=$ constant and pressure $p=$ constant for $t=mt_{s}$ with
$m=0,1,2,....$. So the big bang like singularity is also
regularized. For both the models, we have examined the validity of
null energy condition and strong energy condition. For sine model,
null energy condition is satisfied for $0<\alpha\le 4$ with $k\ge
0$ and strong energy condition is satisfied for for $0<\alpha\le
1$. But for tangent model, null energy condition is satisfied for
$0<\alpha\le 2$ and $k\ge 0$ and strong energy condition is
satisfied for $0<\alpha\le 1$.\\

Finally, we have studied the thermodynamic nature in Multiverse
($n$ number of universes) model. Using the first law of
thermodynamics, we have obtained the total entropy of the
Multiverse for (i) varying $c$ with constant $G$ model, (ii)
varying $G$ with constant $c$ model and (iii) both varying $c$ and
varying $G$ model. Then have shown that the total entropy is
always the same (constant) provided there are some relations
between $n$ number of universes, where $n$ is considered as even
number. We have assumed the speed of light and gravitational
constant of $n/2$ number of universes are growing/diminishing and
their values of other $n/2$ number of universes are
diminishing/growing. So for both varying speed of light and
varying gravitational constant, we have concluded that total
entropy of the Multiverse is constant if the entropies of $n/2$
number of universes are growing/diminishing and entropies of other
$n/2$ number of universes are diminishing/growing.\\\\

{\bf Acknowledgement}: One of the author SN has calculated part of this work during his M. Sc. project in 2019. \\\\


\begin{thebibliography}{99}

\bibitem{Carr} B. Carr, ed., Universe or Multiverse (Cambridge University
Press, Cambridge, UK, 2007).
\bibitem{Gibb} G. W. Gibbons, S. W. Hawking and S. T. C. Siklos, eds., ``Natural
Inflation''. The Very Early Universe (Cambridge University Press,
1983) p. 251.
\bibitem{Vil} A. Vilenkin, Phys. Rev. D 27, 2848 (1983).
\bibitem{Reb} E. Rebhan, Int. J. Mod. Phys. A 32, 1750149 (2017).
\bibitem{Rob} S. J. Robles-Perez, Universe 3, 49 (2017).
\bibitem{Robl} S. Robles-Perez, A. Balcerzak, M. P. Dabrowski and M. Kramer, Phys. Rev. D 95, 083505 (2017).
\bibitem{Mora} J. Morais, M. Bouhmadi-Lopez, M. Kramer and
S. Robles-Perez, Eur. Phys. J. C 78, 240 (2018).
\bibitem{Bou} M. Bouhmadi-Lopez, M. Kramer, J. Morais and S. Robles-Perez, JCAP 1902, 057 (2019).
\bibitem{Roble} S. Robles-Perez, A. Alonso-Serrano, C. Bastos and O. Bertolami, Phys. Lett. B 759, 328 (2016).
\bibitem{Mar} K. Marosek, M. P. Dabrowski and A. Balcerzak, MNRAS 461, 2777 (2016).
\bibitem{Mar2} M. P. Dabrowski and K. Marosek, JCAP 1302, 012
(2013).
\bibitem{Mar3} M. P. Dabrowski, K. Marosek and A. Balcerzak, Memorie della Societa
Astronomica Italiana 85, 44 (2014).
\bibitem{A1} A. Albrecht and J. Magueijo, Phys. Rev. D 59, 043516 (1999).
\bibitem{B1} J. D. Barrow, Phys. Rev. D 59, 043515 (1999).
\bibitem{B2} J. D. Barrow and J. Magueijo, Class. Quant. Grav. 16, 1435 (1999).
\bibitem{Chak1} S. Chakraborty, N. C. Chakraborty and U. Debnath, Int. J. Mod. Phys. D 11, 921 (2002).
\bibitem{Chak2} S. Chakraborty, N. C. Chakraborty and U. Debnath, Int. J. Mod. Phys. A 18, 3315 (2003).
\bibitem{Chak3} S. Chakraborty, N. C. Chakraborty and U. Debnath, Mod. Phys. Lett. A 18, 1549 (2003).
\bibitem{UD2} S. Chakraborty, N. C. Chakraborty and U. Debnath, Int. J. Mod. Phys. D 12, 325 (2003).
\bibitem{Thorne} K. S. Thorne, Astrophys. J. 148 51 (1967).
\bibitem{AM} A. Albrecht and J. Magueijo, Phys. Rev. D 59, 043516 (1999).
\bibitem{Tipler} F. J. Tipler, Phys. Lett. A 64, 8 (1977).
\bibitem{Cald} R. Caldwell, Phys. Lett. B 545, 23 (2002).
\bibitem{Barr} J. D. Barrow, G. J. Galloway and F. J. Tipler, Mon.
Not. R. Astron. Soc. 223, 835 (1986).
\bibitem{Nojiri}  S. Nojiri, S. D. Odintsov and S. Tsujikawa, Phys. Rev. D 71, 063004 (2005).
\bibitem{Dab} M. P. Dabrowski and T. Denkiewicz, Phys. Rev. D 79, 063521 (2009).
\bibitem{Fram} P. H. Frampton, K. J. Ludwick and R. J. Scherrer, Phys. Rev.
D 84, 063003 (2011).

\bibitem{UD4} S. Chakraborty and U. Debnath, Grav. Cosmo. 17, 280 (2011).
\bibitem{UD5} S. Chakraborty and U. Debnath, Int. J. Mod. Phys. D 19, 2071 (2010).
\bibitem{UD6} M. F. Shamir, Adv. High Energy Phys. 2017, 6378904 (2017).
\bibitem{UD7} M. F. Shamir and F. Kanwal, Eur. Phys. J. C 77, 286 (2017).
\bibitem{UD8} M. F. Shamir and A. Komal, Int. J. Geom. Meth. Mod. Phys. 14, 1750169 (2017).
\bibitem{a1} C. B. Collins, E. N. Glass and D. A. Wilkinson. Gen. Rel. Grav. 12, 805 (1980).


\end{thebibliography}
\end{document}